# Maltese Cross anisotropy in Ho$_{0.8}$Lu$_{0.2}$B$_{12}$ antiferromagnetic metal with dynamic charge stripes.

A.L. Khoroshilov[a,b], V.N. Krasnorussky[a], K.M.Krasikov[b,a], A.V. Bogach[a], V.V. Glushkov[a,b], S.V. Demishev[a,c], N.A. Samarin[a], V.V. Voronov[a], N.Yu. Shitsevalova[d], V.B. Filipov[d], S. Gabáni[e], K. Flachbart[e], K. Siemensmeyer[f], S.Yu. Gavrilkin[g], N. E. Sluchanko[a,b,*]

[a]*Prokhorov General Physics Institute of Russian Academy of Sciences, Vavilova 38, 119991 Moscow, Russia*

[b]*Moscow Institute of Physics and Technology (State University), Moscow Region 141700 Russia*

[c]*National Research University Higher School of Economics, Myasnitskaya ulitsa, 20, Moscow 101000, Russia*

[d]*Institute for Problems of Materials Science, NASU, Krzhizhanovsky str., 3, 03142 Kyiv, Ukraine*

[e]*Institute of Experimental Physics SAS, Watsonova 47, 04001 Košice, Slovakia*

[f]*Hahn Meitner Institut Berlin, D 14109 Berlin, Germany*

[g]*Lebedev Physical Institute of Russian Academy of Sciences, Leninskiy Prospekt 53, 119991, Moscow, Russia*

*E-mail: nes@lt.gpi.ru

The model strongly correlated electron system Ho$_{0.8}$Lu$_{0.2}$B$_{12}$ which demonstrates a cooperative Jahn-Teller instability of the boron sub-lattice in combination with rattling modes of Ho(Lu) ions, dynamic charge stripes and unusual antiferromagnetic (AF) ground state has been studied in detail at low temperatures by magnetoresistance ($\Delta\rho/\rho$), magnetization and heat capacity measurements. Based on received results it turns out that the angular *H-φ-T* magnetic phase diagrams of this non-equilibrium AF metal can be reconstructed in the form of a "Maltese cross". The dramatic AF ground state symmetry lowering of this dodecaboride with *fcc* crystal structure can be attributed to the redistribution of conduction electrons which leave the RKKY oscillations of the electron spin density to participate in the dynamic charge stripes providing with extraordinary changes in the indirect exchange interaction between magnetic moments of $Ho^{3+}$ ions and resulting in the emergence of a number of various magnetic phases. It is also shown that the two main contributions to magnetoresistance in the complex AF phase, the (*i*) positive linear on magnetic field and the (*ii*) negative quadratic $-\Delta\rho/\rho\sim H^2$ component can be separated and analyzed quantitatively, correspondingly, in terms of charge carrier scattering on spin density wave (*5d*) component of the magnetic structure and on local *4f-5d* spin fluctuations of holmium sites.

# I. INTRODUCTION.

The complexity of strongly correlated electron systems (SCES) is a subject of active debates, and numerous investigations have been carried out to clarify its nature [see e.g. [1-2]). In recent years it was demonstrated that at least some of SCES are spatially inhomogeneous materials and that their electronic complexity arises from charge, spin, lattice and orbital degrees of freedom which act simultaneously, leading to giant responses at small perturbations [1-5]. Moreover, when several metallic and insulating phases compete, it creates the potential for novel behavior and practical applications, and well-known examples of it are the Mn oxides called manganites [1, 6-12], high temperature superconducting cuprates [1-4, 13-16], iron-based pnictides and chalcogenides [4-5, 17-20], etc. Among the widely discussed issues in these materials are very complicated phase diagrams with various magnetic phases and ground states in combination with diverse mechanisms responsible for their competition and stabilization [1-20]. According to conclusions of ref. 7, the ground states diversity and mixed-phase tendencies have two origins: (*i*) electronic phase separation between phases with different densities that leads to nanometer scale coexisting clusters, and (*ii*) disorder-induced phase separation with percolative characteristics between equal-density phases which is driven by disorder near first-order metal-insulator transitions. In this regard, static and dynamic charge stripes also refer to factors of diversity which can be from the strong-coupling perspective considered as a real-space pattern of micro phase separation [21].

Taking into account that the search of mechanisms responsible for the SCES complexity is significantly hampered by their complex multicomponent chemical composition and complicated unit cells with a low symmetry crystal structure, it is promising to use instead model compounds with charge stripes and magnetic or, superconducting ground states in combination with a simple cubic crystals structure. Very recently, fluctuating charge stripes have been detected in non-magnetic face centered cubic ($fcc$, see Fig.1a-c for details) dodecaboride $LuB_{12}$ [22-23] and their origin was explained in terms of a cooperative dynamic Jahn-Teller effect in boron $B_{12}$ cubooctahedra. Moreover, it was shown that $Ho_xLu_{1-x}B_{12}$ are antiferromagnetic disordered metals with emergence of nanosize AF domains in the paramagnetic (P) state far above the Neel temperature ($T_N$), and with a strong concurrence of large positive and negative effects in magnetoresistance [24] similar to those detected in manganites. Additionally, strong magnetotransport anisotropy was observed in the paramagnetic state of $Ho_{0.8}Lu_{0.2}B_{12}$ and it was interpreted in terms of charge carrier scattering on dynamic charge stripes (DCS) oriented along <110> direction in the $fcc$ lattice [25].

In order to utilize these model compounds and to shed more light on the origin of various magnetic phases in AF metals with dynamic charge stripes, in this study detailed investigation was undertaken to characterize both the anisotropy of magnetic ground state and the phase transitions in $Ho_{0.8}Lu_{0.2}B_{12}$ below $T_N \approx 5.7$ K [24-25]. It was observed that in this AF metal with indirect exchange between localized magnetic moments of holmium ions through conduction electrons (RKKY-mechanism), the occurrence of dynamic charge stripes leads to the formation of a strongly anisotropic phase diagram with a large variety of different magnetic phases. It is shown that both a Maltese cross anisotropy of charge transport in steady magnetic field and the same symmetry arrangement of phase boundaries in $H$-$\varphi$-$T_0$ magnetic phase diagrams of the AF state allow us to distinguish three sectors located in the vicinity of three main directions: (*i*) along stripes ($\boldsymbol{H}$//[110]), (*ii*) transverse to stripes ($\boldsymbol{H}$//[001]) and (*iii*) in the direction of the magnetic structure ($\boldsymbol{H}$//[111]). The analysis of positive and negative MR components is also carried out providing information about mechanisms of charge carriers scattering in the spatially inhomogeneous AF state with an electronic nanometer size phase separation.

## II. EXPERIMENTAL METHODS.

Precise measurements of resistivity $\rho$ and transverse magnetoresistance $\Delta\rho/\rho$ in the AF and P phases of $Ho_{0.8}Lu_{0.2}B_{12}$ at temperatures in the range 1.9-20 K were performed in an external magnetic field $\boldsymbol{H}$ up to 80 kOe for different orientations relative to the main crystallographic directions. The sample holder allowed sample rotation with a step-by-step change of the angle $\varphi = \boldsymbol{H}^{\wedge}\boldsymbol{n}$ between the normal $\boldsymbol{n}$ to the surface (110) of the crystal and the direction of the external steady magnetic field $\boldsymbol{H}$ in the range $\varphi = 0 \div 360°$ with a step $\Delta\varphi = 1.8°$ (see Fig. 1d). The resistivity was measured by the standard 4-terminal direct current technique with commutation of measuring current $\boldsymbol{I}$//[1-10] [25]. High precision of the temperature stabilizations ($\Delta$T=0.001 K) was provided by the temperature controller, model TC1.5/300 (Cryotel Ltd). High accuracy control of the steady magnetic field was achieved by the power supply SMPS-200 (Cryotel Ltd). To confirm the location of phase boundaries on the $H$-$\varphi$-$T_0$ diagrams, commercial Quantum Design products MPMS-5 and PPMS-9 were used at low temperatures 1.9-20 K for measurements of magnetization and heat capacity in magnetic field up to 50 kOe and 90 kOe, correspondingly for $\boldsymbol{H}$//[001]), $\boldsymbol{H}$//[110] and $\boldsymbol{H}$//[111]. High quality single-domain crystals of $Ho_{0.8}Lu_{0.2}B_{12}$, similar to those studied in the paramagnetic state in [25], were grown by the technique of vertical induction non-crucible melting in an inert gas atmosphere [26]. The magnetic field corrections due to the demagnetization of samples studied were estimated and did not exceed 3%.

## III. EXPERIMENTAL RESULTS.

**III.1. Resistivity, specific heat and susceptibility behavior near $T_N$.** Fig. 2 shows the temperature dependences of ~~the~~ resistivity, magnetic susceptibility (H = 100 Oe), and heat capacity of $Ho_{0.8}Lu_{0.2}B_{12}$ in a wide vicinity of the antiferromagnet (AF) – paramagnet (P) phase transition. As can be seen from Fig. 2, the susceptibility curve $\chi(T)$ is typical for antiferromagnets with a sharp peak at ~~the~~ Neel temperature $T_N = 5.7$ K. Near $T_N$, the temperature dependences of ~~the~~ resistivity and heat capacity also show strong anomalies corresponding to sharp step-like growth during the transition to the AF phase. It is worth noting the formation of a wide minimum ($T_{min} \sim 11$ K) and the subsequent noticeable increase of resistivity and specific heat with decreasing temperature before transition to the Neel state. Temperature lowering in the magnetically ordered phase of $Ho_{0.8}Lu_{0.2}B_{12}$ at T < 5.5 K is accompanied by a significant decrease in resistivity and specific heat and it induces only small changes in the magnetic susceptibility.

**III.2. Magnetoresistance in magnetic field along three principal directions of the cubic structure.** Fig. 3 shows the results of measurements of the magnetoresistance (MR) $\Delta\rho/\rho=(\rho(H)-\rho(0))/\rho$ (0) of $Ho_{0.8}Lu_{0.2}B_{12}$ samples in the temperature range 1.9–6.6 K in magnetic fields up to 80 kOe for three principal directions in the cubic structure $\mathbf{H} \parallel [001]$, $\mathbf{H} \parallel [110]$ and $\mathbf{H} \parallel [111]$ (panels a-c, respectively, arrows show several AF-P transitions and orientation magnetic phase transitions in the AF state). Among the main features of MR, the following should be noted: (*i*) a significant positive magnetoresistance (pMR) in the antiferromagnetic phase, whose amplitude increases with decreasing temperature, (*ii*) ~~a~~ near-linear changes of MR in the low field region for directions $\mathbf{H} \parallel [001]$, $\mathbf{H} \parallel [110]$, (*iii*) the presence of two linear pMR intervals for orientation $\mathbf{H} \parallel [111]$ in ranges of 0–10 kOe and 15–30 kOe, respectively, with significantly different slopes, and (*iv*) an abrupt decrease of MR with increasing H in the vicinity of the AF-P phase transition. The pMR maximum (~ 98%) in the AF state is achieved in the field direction **H** ∥ [001] at T = 1.9 K, while the highest absolute values of negative magnetoresistance (nMR) ~ 35% are observed in a wide range of fields H ≤ 80 kOe in the paramagnetic phase for **H** ∥ [111] at T = 5.4 K (see, also, [25]). We also note a presence of different anomalies on $\Delta\rho/\rho=f(H, T_0)$ curves associated with orientation magnetic phase transitions in the AF state. To reconstruct the magnetic H – T phase diagram and to attribute the features on magnetoresistance dependencies with phase transitions, we measured also the magnetic and thermal properties of $Ho_{0.8}Lu_{0.2}B_{12}$ single crystals.

**III.3. Heat capacity in magnetic field.** Fig. 4 shows the temperature dependences of specific heat in magnetic fields up to 70 kOe for three principal directions in the cubic structure **H** ∥ [001], **H** ∥ [110] and **H** ∥ [111], arrows indicate magnetic phase transitions. It is easy to see that there is a minimum in the paramagnetic region on $C(T,H_0)$ curves that shifts with increasing magnetic field upwards in temperature. At the same time, the growth of external magnetic field initiates the appearance of a low-temperature maximum in the P-phase. On the contrary, a large amplitude step-like behaviour that corresponds to AF-P transition shifts with increasing H downward, demonstrating the typical tendency to suppress the AF magnetic order state by strong external magnetic field. Separately, we note a number of features in the form of sharp peaks and additional steps below $T_N$, corresponding to orientation magnetic phase transitions (see, for example, Fig. 4b).

**III.4. Magnetization in the vicinity of magnetic phase transitions.** Fig. 5 shows the field dependences of magnetization per holmium ion $M(H,T_0)$ for three principal field directions up to 50 kOe in the temperature range of 1.9–6.6 K (panels a – c), as well as derivatives $dM/dH=f(H, T_0)$ obtained by numerical differentiation (panels d – f). For clarity, the region of magnetic phase transitions in the AF phase is shown on an enlarged scale in Fig.6a – c, where the orientation phase transitions are marked by arrows. As can be seen from Figs. 5a-c, the AF-phase transition only slightly changes the absolute magnetization values in the entire range of H. At the same time, details of anomalies on dM/dH = f(H) curves associated with magnetic phase transitions in the range of 5-25 kOe are noticeable different for various directions of the external magnetic field. Indeed, the narrowest features corresponding to sharp phase transitions are recorded in the direction **H** ∥ [111]. On the contrary, for **H** ∥ [001], are the low-temperature anomalies ~~are~~ significantly broadened (see Fig.6a-c). It is worth noting also that the magnetic field interval where the orientation phase transitions are observed for **H** ∥ [110] at temperatures of 1.9 - 2.1 K turns out to be noticeably wider than for two other directions. In particular, in the orientation **H** ∥ [110], the feature is reliably recorded at $H_0 \approx 22$ kOe, whereas for **H** ∥ [001] and **H** ∥ [111], the upper limit of orientation magnetic transitions according to ~~the~~ magnetization data is located no higher than at H ≈ 18 kOe (Fig. 6a − c). Fig. 7 presents a comparison of ~~the~~ features on ~~the~~ curves corresponding to the orientation phase transitions for different magnetic field directions (panels a, d and g) with heat capacity data (panels Fig. 7b, e, h) and obtained by numerical differentiation of the $\Delta\rho/\rho = f(H, T_0)$ curves at T = 2.1 K for field directions **H** ∥ [001], **H** ∥ [110], and **H** ∥ [111] (transition fields are shown by vertical dashed lines). As can be seen from Fig. 7, the accuracy of the analysis of critical fields and, hence, the precision of the magnetic H-T phase diagram can be significantly increased by using $\Delta\rho/\rho = f(H, T_0)$ derivatives, in addition to the initial MR data and thermodynamic dependencies of C(H,T) and M(H,T). Moreover, the analysis of MR derivatives allows us to reveal a phase transition in field H = 52 kOe for the direction **H** ∥ [001] which is not evident on the curves of magnetoresistance, heat capacity and magnetic susceptibility (see the inset in Fig.7c).

### III.5. Field dependences of magnetoresistance for various directions of magnetic field.

The application of a rotating sample holder rotating in ~~a~~ magnetic field (see its schematic view in fig.1d)) allowed us to perform measurements of magnetoresistance for different field directions in a wide range of angles $\varphi = \mathbf{H}\hat{\ }\mathbf{n}$ between vector $\mathbf{H}$ and the normal to the sample surface $\mathbf{n}\|[001]$ for a detailed study of charge transport anisotropy and magnetic phase diagram. The results of the magnetoresistance measurements in fields up to 80 kOe at temperature $T_0 = 2.1$ K which were undertaken in a wide range of angles $\varphi=0\text{-}90°$ are shown in Fig. 8 (magnetic phase transitions are indicated by arrows). In supplement [27] also MR curves measured at $T_0 = 4.2$ K are shown. Figs.8a and 8b show $\Delta\rho/\rho=f(H)$ dependences for angles in the range from $\varphi=0°$ (direction $\mathbf{H} \| [001]$) to $\varphi=54°$ ($\mathbf{H} \| [111]$) and in the interval $\varphi=54\text{-}90°$, correspondingly ($\varphi=90°$ corresponds to the orientation of $\mathbf{H} \| [110]$).

Describing the evolution of ~~the~~ field dependences of MR ~~recorded~~ in experiments with ~~a~~ sample rotating in magnetic field, it is convenient to distinguish three main sets of curves corresponding to angle intervals $\Delta\varphi_{001} = 0\text{-}36°$ (sector in vicinity of $\mathbf{H} \| [001]$), $\Delta\varphi_{111} = 36\text{-}76°$ (area of direction $\mathbf{H} \| [111]$) and $\Delta\varphi_{110} = 76\text{-}90°$ (neighborhood of the $\mathbf{H} \| [110]$). We emphasize that the field dependencies in each of these sectors have their own unique set of features that transform into each other when the direction of the external magnetic field changes. Thus, $\Delta\rho/\rho=f(H, T=2.1$ K$)$ curves near the direction $\mathbf{H} \| [001]$ demonstrate low-field features in the range of 15–18 kOe, allowing also to detect the AF-P phase transition at $H_N \approx 64$ kOe, which is preceded by a wide MR decrease with increasing H.

When $\mathbf{H}$ deviates from the [001] direction, an additional feature of magnetoresitance appears slightly below the AF-P transition (see the curves for $\varphi=5°$, 18° and 27° in Fig. 8a). As a result a double step singularity emerges below $H_N$ consisting of a small high-field step and a large amplitude low-field anomaly. The high-field step increases with increasing ~~angle~~ $\varphi$ up to 36°, while the feature itself shifts slightly toward low magnetic fields. On the contrary, the low-field large step of the double transition decreases, shifts to low H values and with increasing angle practically disappears near the direction $\mathbf{H} \| [112]$ ($\varphi=36°$). Worth noting is also an abrupt step-like anomaly of MR arising on curves for $\varphi \geq 36°$ in magnetic fields H $\leq$ 32 kOe, which shifts sharply with increasing φ in the interval 36-40°, and performs a transition from one slope behavior to another (see, for example, the curve φ =40° in Fig. 8a, as well as [27]).

In the range of angles φ =45-63° in vicinity of direction **H** ∥ [111], the small-field transition between two linear asymptotes of MR splits into two (see e.g. curves for φ = 45°, 50° and 54° in the range of 12-15 kOe in Fig. 8a). As φ changes from 54° to 63°, these small-field MR features change in reverse order and the double-step anomaly disappears. But, in the range φ =58 -72° another double-step structure arises on $\Delta\rho/\rho$ =f(H, $T_0$) curves in fields of 21 -40 kOe (Fig. 8b). A further increase of φ in the interval φ > 72° leads to a modification of the right step-like $\Delta\rho/\rho$ =f(H, $T_0$) anomaly to a form similar to that observed for **H** ∥ [001]. At the same time, the left step-like singularity discussed above is in the angular interval φ = 81-90° transformed into a transitional region between two linear parts on MR curves which are very close in slope. Moreover, on curves for directions around **H** ∥ [110] in magnetic fields immediately below the AF-P transition, an additional feature arises in the form of a small amplitude maximum (see curves for angles φ =76-90° in the field interval of 49-55 kOe in Fig. 8b). For comparison, in [27] a similar set of MR field dependences at $T_0$ = 4.2 K is shown. Note that with temperature increase from 2.1 K to 4.2 K, the general view of the picture in the neighborhood of each principal H directions **H** ∥ [001], **H** ∥ [110], and **H** ∥ [111] does not change, however, at $T_0$=4.2 K the amplitude of the positive contribution to magnetoresistance decreases significantly (by a factor of 3 compared with MR for $T_0$=2.1 K) and, at the same time, the intervals of magnetic field where these MR anomalies observed become narrowed significantly (see [27]).

**III.6. Angular dependences of resistivity.** Additional information on the location of ~~the~~ phase boundaries between ~~these~~ selected sectors of $\Delta\varphi_{001}$, $\Delta\varphi_{110}$ and $\Delta\varphi_{111}$ in a wide neighborhood of three principal crystallographic directions can be obtained from angular dependences of resistivity. For example, Fig. 9 shows the full range (φ = 0- 360°) dependencies in magnetic field up to 80 kOe, measured at T = 2.1 K. In low fields H <12 kOe, the maximum values of pMR in the AF phase are observed for **H** ∥ <112>, while the minimum values are recorded along **H** ∥ <110> diagonals. With increasing field in the range of 10-15 kOe along **H**∥<111> a sharp minimum forms (see the curve for $H_0$ = 12 kOe in Fig.9) which significantly broadens at $H_0$=15 kOe. In this case, the sharp changes of resistivity observed on the borders of the minimum interval obviously correspond to phase transitions in the AF state. In magnetic fields 15-50 kOe, along with the minimum values near **H** ∥ <111>, the highest resistivity is observed in a wide neighbourhood of directions **H** ∥ <001> and **H** ∥ <110> (see fig. 9). The switching between ρ(φ) maxima and minima occurs abruptly near **H** ∥ <112>. As the field increases in the range of 40–80 kOe, the feature of the form of a wide "plateau" in the vicinity of **H**∥<001> (Fig.9b) transforms into a bell-shaped maximum, tapering in the range of 55 - 62 kOe (Fig.9c) and then broadening in the paramagnetic phase when H increases (Fig.9d). Note also that the maximum in the neighborhood of **H** ∥ <110> decreases sharply and is transformed to a minimum in the field interval 55–60 kOe (Fig. 9d). In the P-phase, on curves in the range of 50–130°, a combined anomaly is observed which consists of three shallow minima near directions **H** ∥ <111> and **H** ∥<110> (see fig. 9d, and also [25]).

The angular dependences of resistivity ρ(φ, H) measured at 4.2 K, which demonstrate similar behavior, are given in [27]. Note also that the "resistivity plateau" at 4.2 K in the neighborhood of **H**∥ <001> and **H** ∥ <110> becomes much wider due to the narrowing of angular intervals with almost constant resistivity in the vicinity of **H** ∥ <111>. Moreover, for $T_0$ = 4.2 K, the position of the abrupt changes of ρ(φ) no longer correspond to crystallographic directions **H** ∥ <112>, as it was observed at $T_0$ = 2.1 K (see fig. 11-12). It should be emphasized that the experimental data presented in this section (see also [27]) indicate a very strong anisotropy of both the magnetic phase diagram of $Ho_{0.8}Lu_{0.2}B_{12}$ and the charge carrier scattering in this model strongly correlated electron system.

## IV. DISCUSSION.

**IV.1. H-T magnetic phase diagram for directions H ∥ [001], H ∥ [110] и H ∥ [111].** The results obtained in the studies of magnetoresistance, heat capacity, and magnetization (Fig. 2-7) allow us to reconstruct the magnetic phase diagrams of the $Ho_{0.8}Lu_{0.2}B_{12}$ antiferromagnet for magnetic field directed along the three principal cubic directions, - **H** ∥ [001], **H** ∥ [110] and **H** ∥ [111] (see Fig. 10). It is easy to see that the phase boundaries, between various magnetic phases in the AF state and the location of the AF-P transition, agree very well with each other.

When analyzing the anisotropy, there is only a small difference in the position of the AF-P phase boundary for **H** ∥ [001] ($H_N$ (T = 1.9 K) ≈ 64 kOe) and for two other directions ($H_N$ (T = 1.9 K) ≈ 61-62 kOe) (see Fig. 10). On the contrary, in the AF phase itself, both the type and the location of the boundaries differ significantly depending on the orientation of the magnetic field. For example, for **H** ∥ [001] in addition to the low-field branch on H-T diagram, there is a new phase boundary falling from 55 kOe to 15 kOe, and a triple point at $T_{int1}$ = 3.3 K (Fig. 10a, the location of this transition was also refined by the temperature dependences of resistivity, see [27]). Moreover, as can be seen from fig. 10, especially for **H** ∥ [110] (panel b), the low-field branch of the phase diagram splits into two parts below 4.2 K, and the additional phase between them is practically not resolved in the orientation **H** ∥ [100]. Among the features of the phase diagram, we also note that in intermediate fields 15-55 kOe for **H** ∥ [110] an additional phase boundary is recorded (Fig. 10b, see also [27]), whereas for **H** ∥ [111] in the range of 3.2–3.4 K an intersection of two low-field branches is observed (Fig. 10c). Note that the intersection point of these two branches $T_{int2}$ ≈ 3.3 K coincides with $T_{int1}$ detected for **H** ∥ [001] (indicated by arrows in fig.10a and 10c).

**IV.2. Diagrams of charge carrier scattering and phase transitions.** Data presented above indicate evidence of strong anisotropy of charge transport, magnetic and thermal properties in the AF phase of $Ho_{0.8}Lu_{0.2}B_{12}$. Therefore, the MR data obtained at fixed temperatures (vertical sections at 2.1 K, 4.2 K and 6.5 K in the phase diagram in Fig.10), or at fixed magnetic field

(horizontal section at 25 kOe in Fig. 10) in a 3D plot allow to see the evolution of transverse magnetoresistance with temperature lowering and with the magnitude and direction of applied magnetic field. These results obtained for the AF state at $T_0 = 2.1$ K and 4.2 K are presented in Fig. 11a-11b, respectively. Fig. 11c demonstrates a 3D plot of MR for $H_0 = 25$ kOe and, for comparison, Fig. 11d shows the MR anisotropy in the paramagnetic phase ($T_0 = 6.5$ K, see also [25]). In fig. 12a-12c for visualization of phase boundaries in (H, φ, T) planes, we also present the ~~color~~ projections of MR data onto the (1-10) plane. It is easy to see that in polar plot of the transverse MR of the single-domain $Ho_{0.8}Lu_{0.2}B_{12}$ crystal with the current direction [1-10] (**H** // (1-10), see Fig.11) the anisotropic phase diagrams have the shape of a Maltese cross with lobes, which are determined by the width of the sectors $\Delta\varphi_{001}$ and $\Delta\varphi_{110}$.

We also emphasize that the sharp radial phase boundaries of the $\Delta\varphi_{001}$, $\Delta\varphi_{110}$ and $\Delta\varphi_{111}$ sectors, which form a Maltese cross, imply that the magnetic phases corresponding to these sectors differ significantly from each other in their magnetic structure. At the same time, the set of phases (indicated by Roman numerals in Figs 12a-12c), which are observed in each of these three angular segments listed above, is different. Such a complicated and low-symmetry magnetic H-φ-T diagram is unusual and unexpected for a rare earth dodecaboride with a *fcc* crystal structure. It is obviously a consequence of additional strong charge carrier scattering which modifies dramatically the magnetic interaction of 4*f* moments of Ho-ions through conduction electrons (RKKY mechanism).

When discussing the nature of the observed dramatic symmetry lowering, we note that significant MR changes are observed already in the paramagnetic phase (see e.g. Fig.11d for $T_0 = 6.5$ K) as shaping regions with maximum negative values of nMR (~ 30%) in a wide neighborhood of diagonal directions **H** || [110] and **H** || [111] together with the sector of minimal nMR effect (1-3%) for **H** || [100]. It is easy to see that similar anisotropy is also maintained at temperatures of 2.1 K and 4.2 K (see Figs. 11a and 11b, respectively). Thus, for all three values of temperature $T_0$=2.1 K, 4.2 K and 6.5 K, there is a general tendency to form a sector in high magnetic fields with maximum values of MR in the direction **H** ∥ [001] in combination with regions which are characterized by minimal MR and located near the diagonals **H** ∥ [110] and **H** ∥ [111] (see fig. 8-9, 11-12, as well as [27]). We argue, that numerous magnetic phases observed below $T_N$ and a lot of phase transitions in the H-φ-T diagram in the AF state are the results of two factors: the effect leading to uniaxial scattering anisotropy in the paramagnetic phase (dynamic charge stripes in the matrix of dodecaborides, see below) and the magnetic ordering due to indirect AF exchange interaction through the conduction electrons (RKKY-mechanism).

It was shown previously [25] that the magnetoresistance in the paramagnetic phase of $Ho_{0.8}Lu_{0.2}B_{12}$ can be represented as a sum of (*i*) isotropic nMR observed at T < 10 K which is associated with charge carrier scattering on magnetic clusters of $Ho^{3+}$-ions (nanometer size domains with AF short-range order) and (*ii*) anisotropic pMR due to the cooperative dynamic Jahn-Teller effect (JT) in the boron sub-lattice. Indeed, it was found [22-23] that the cooperative dynamic JT-effect on the $B_{12}$ molecules produces rattling modes of heavy R-ions causing to dynamic changes in *5d-2p* hybridization ("modulation" of the conduction band) and resulting into emergence of dynamic charge stripes in the *fcc* metallic matrix of dodecaborides. It was detected [23] that in $LuB_{12}$ stripes are induced along [110] directions of the *fcc* crystal structure. Recent studies of the dynamic conductivity of $LuB_{12}$ [28] found that already at room temperature most (over 70%) of the conduction electrons are involved in formation of a collective mode (over-damped oscillator) in these metals with a band width of about 1.6 eV [29]. Similar effects of charge transport anisotropy in semiconducting solid solutions $Tm_{0.19}Yb_{0.81}B_{12}$ and formation of charge stripes already at room temperature have been reported in [30]. Thus, it should be expected that upon transition to the disordered cage-glass state (T*~ 60 K, [31-32]), which is well above the transition to the magnetically ordered phase in $Ho_{0.8}Lu_{0.2}B_{12}$ ($T_N \approx 5.7$ K ), the crystal structure of the dodecaboride becomes distorted because of JT instability of the boron cage and results into the appearance of dynamic charge stripes. These fluctuating charges along [110] lead to a strong anisotropic scattering of charge carriers in a magnetic field and, as a result, to the appearance of an anisotropic positive contribution to ~~the~~ magnetoresistance. In such a case of the quantum motion of charge density in the conduction band (alternating current of a quantum-mechanical nature) two directions in the *fcc* lattice (*i*) along (**H** // [110]) and transverse to the stripes (**H** // [001]) become distinguished, leading to the formation of the Maltese cross structure both in the transverse MR data and in the H-φ-T magnetic diagram of $Ho_xLu_{1-x}B_{12}$ antiferromagnetic solid solutions. It should be noted that in this picture the neighborhood of the [111] direction is also highlighted. In this direction the moments of holmium in $Ho_xLu_{1-x}B_{12}$ dodecaborides are oriented in an incommensurate multiple-q AF-phase with basic propagation vector $q_{AF}$ // (111) [33]. As a result, since the effect of transverse magnetoresistance in a steady external magnetic field is completely determined by the nature of charge carrier scattering in ~~the~~ crystals, the magnetic phase diagrams in polar plots (φ, H) and (φ, T) present scattering patterns of conduction electrons on the antiferromagnetic structure in the presence of dynamic charge stripes along the [110] axis.

**IV.3. Scaling of the magnetoresistance in the AF-state.** It can be seen that the MR field dependences in the AF phase (Fig. 3, $T_0 = 1.9$–5.4 K) for all given field directions demonstrate extended linear sections $\Delta\rho/\rho(H)$ in moderate magnetic fields. In Fig. 13, the same set of field dependences of ~~the~~ magnetoresistance is shown for three magnetic field directions in coordinates $\Delta\rho/\rho = f$ (H/T). Obviously, the indicated linear parts of the curves practically coincide in the representation we used. In particular, for **H** ∥ [001] scaling is observed in the range H/T≤ 20 kOe/K, for **H** ∥ [110] up to 16 kOe/K, and for **H** ∥ [111] there are two similar intervals, - H/T≤ 6 kOe/K and the range 7-20 kOe/K, where the MR curves are well approximated by linear dependencies (indicated by dotted straight lines in Fig. 13). The scaling of $\Delta\rho/\rho = f(H/T)$ curves in a wide range of temperatures for different field directions in the AF phase indicates the presence of a linear contribution to magnetoresistance in the AF state of

$Ho_{0.8}Lu_{0.2}B_{12}$, whose amplitude scales with temperature. Below we present a detailed analysis of these results obtained with the separation of the main contributions to magnetoresistance.

**IV.4. Analysis of MR components in the AF state.** In the framework of approach developed in [25, 34], in the AF phase of $Ho_xLu_{1-x}B_{12}$ in the moderate magnetic fields, MR was proposed to be considered as a sum of the negative quadratic in the magnetic field and the positive linear contributions

$$\frac{\Delta\rho}{\rho_0}(H, T_0) \approx -\frac{B(T_0)}{2}H^2 + A(T_0)H \qquad (1),$$

where $B(T_0)$, $A(T_0)$ are coefficients depending on temperature and magnetic field direction. The linear positive contribution to pMR is traditionally associated with carrier scattering on a spin density waves (SDW) (see e.g. [35−37]). Such behavior is typical, for example, for the itinerant antiferromagnetic phase in chromium, in the absence of Fe and Co magnetic impurities [38]. On the other hand, the negative quadratic component in small magnetic fields nMR in non-Kondo-type systems with metallic conductivity and localized magnetic moments is considered to be induced in the presence of local spin fluctuations resulting to either the emergence of spin-polarization of conduction electrons or to carriers scattering on nanometer size magnetic clusters [24,35-39]. Indeed, the loosely bounded state of rare earth ions in the dodecaboride matrix in combination with the transition into the cage-glass state of $RB_{12}$ at liquid nitrogen temperatures [32], and the appearance of disorder in the arrangement of rare earth ions (random off-site location of $Ho^{3+}$-ions inside the $B_{24}$ truncated cubooctahedron, see fig. 1b) are the factors which result to formation of magnetic nanosize clusters in studied compounds. It was shown in [24,34] that the formula which is similar to one obtained in the Yosida model [40] can be used to estimate the nMR value in P and AF phases of $Ho_xLu_{1-x}B_{12}$ with x≤0.5

$$-\Delta\rho/\rho \sim \boldsymbol{M}_{loc}{}^2 \qquad (2),$$

which in small fields takes the form $-\Delta\rho/\rho \sim \chi_{loc}^2 H^2$ ($M_{loc}$ and $\chi_{loc}$ – local magnetization and susceptibility), and $M_{loc}$ is well described by the Langevin function $M_{loc} \sim L(\alpha)=cth(\alpha)-1/\alpha$ (where $\alpha = \mu_{eff}H/k_B T$ , $k_B$ is the Boltzmann constant and $\mu_{eff}$ the effective magnetic moment of the magnetic nanodomain). Obviously, the weak field approximation is valid only up to 30–40 kOe [24–25]; however, it was established in [25] that relation (2) for the composition $Ho_{0.8}Lu_{0.2}B_{12}$ in the paramagnetic phase up to 80 kOe can be used to estimate lower limit of nMR contribution with an accuracy of no worse than 20%. Taking into account that we are only interested in relative changes of the amplitude of negative and positive contributions to MR in the AF state, we use relation (1) to estimate the charge transport anisotropy and to shed more light on the mechanisms and details of the change of scattering regimes with temperature and magnetic field. Taking into account that the coefficients $A(T_0)$ and $B(T_0)$, which characterize the two abovementioned carrier scattering mechanisms, are changing within phase transitions depending on temperature and both the magnitude and direction of magnetic field, for the MR derivative we have:

$$D_i = \left[\partial\left(\frac{\Delta\rho}{\rho}(H,T_0,\varphi_0)\right)/\partial H\right]_{H\in\Delta H_i} = -B_i(T_0,\varphi_0)H + A_i(T_0,\varphi_0) \qquad (3).$$

Fig.14 shows several examples of the derivative of resistivity by linear dependences (3) (see also Fig. 7c, f, i and [27]). As can be seen from Figure 14, on d$\rho$/dH=f(H) curves we can distinguish three intervals $\Delta H_L$, $\Delta H_M$ and $\Delta H_H$ with close to linear behavior of the derivatives, each of them is characterized by a set of parameters $A(T_0, \varphi_0)$ and $B(T_0, \varphi_0)$. For **H** || [001] and **H** || [110] directions in the $\Delta H_L$ and $\Delta H_M$ intervals separated by phase transitions in low magnetic field, these coefficients characterizing the nMR and pMR contributions coincide within the experimental error, $A_L \approx A_M$ and $B_L \approx B_M$ (see also figs.15-16). On the contrary, for **H** || [111], the parts of the straight lines d$\rho$/dH in Fig. 14b, separated by a double phase transition in the interval 10–20 kOe, correspond to significantly different coefficients $A_L >> A_M$. Moreover, for **H** || [111] on the dependences no linear behavior does is observed in the interval

ΔH$_H$ just below the AF-P transition (see Fig. 14b), which, on the contrary, is reliably recorded for the directions **H** ∥ [001] and **H** ∥ [110] (fig.14a). The analysis of the A(T$_0$,φ$_0$) and B(T$_0$,φ$_0$) behavior obtained within the framework of relation (3) is presented in Fig. 15 and Fig. 16. Figure 15 shows the data for three directions of the magnetic field **H** ∥ [001] (φ$_0$ = 0, direction transverse stripes), **H** ∥ [110] (φ$_0$ = 90°, direction along the stripes) and **H** ∥ [111] (φ$_0$ ≈ 54°, the direction along the magnetic structure) which correspond to temperature dependences of the coefficients A(T$_0$,φ$_0$) (panel a) and B(T$_0$,φ$_0$) (panel b), found for each of three intervals ΔH$_L$, ΔH$_M$ and ΔH$_H$. It can be clearly discerned from Fig. 15a, that the splitting near T∼ 5K of the A(T) dependence for **H** ∥ [111] into two branches A$_L$(T) and A$_M$(T) leads to the appearance of strong charge transport anisotropy and to the formation of H-φ-T diagrams in the form of a Maltese cross (see Fig. 12). As a result, below 5K in the interval ΔH$_M$ of the magnetic field, the slope of Δρ/ρ(H) close to linear for **H** ∥ [111] decreases by a factor of 3-5 compared with the directions **H** ∥ [110] and **H** ∥ [001]. At the same time, a significant anisotropy of the coefficients A$_H$(T) and B$_H$(T) appears in the interval ΔH$_H$ immediately below the AF-P transition (see Fig. 15). We should also note a sharp change in anisotropy of the coefficients B$_L$(T) and B$_M$(T) at temperatures in the vicinity of 5K and near T$_{inv1.2}$ ∼ 3.3 K (see. Fig. 15b). Indeed, above 5K B$_{L,M}$(T) for **H** ∥ [001] is 2-5 times higher than the values for the directions **H** ∥ [110] and **H** ∥ [111], whereas it becomes the smallest below 5K. We emphasize that these changes occur in the vicinity of the intersection of branches in the H − T phase diagram (see Fig. 10) and indicate a different magnetic structure of the magnetically ordered phases in Ho$_{0.8}$Lu$_{0.2}$B$_{12}$. Moreover, since the coefficient B of nMR is a characteristic of local spin fluctuations, it can be assumed that the above mentioned double change of the anisotropy sign B$_{L,M}$(T) (*i*) at 5K and (*ii*) near T$_{inv1.2}$∼3.3 K is associated with a rotation of the preferred direction of *4f-5d* spin fluctuations on Ho$^{3+}$ ions in combination with the magnetic charge carrier scattering caused by them.

Thus, it may be suggested that above 5K, the predominant direction of spin fluctuations on Ho$^{3+}$ ions is the [001] axis, whereas in the interval 3.5–5K, the nMR effect determines the on-site scattering in the [110] and [111] directions (see Fig. 15b). Moreover, taking into account that in the AF phase, the coefficient A in pMR is a measure of charge carrier scattering on a spin density waves, it can be assumed that near 5 K the structure of SDW also changes dramatically. As a result, as H increases up to the field interval $\Delta H_M$, the amplitude of the SDW increases with external magnetic field [41,42]. Thus, this magnetic structure of itinerant electrons manifests itself in the interval of angles in the vicinity of [111] direction, leading to ~~the~~ lowest values of the coefficient A which correspond to the weakest scattering of charge carriers on SDW (fig.15a-16a, s). As can be seen from Figs 16a and 16b, the scattering of conduction electrons on both SDW and $4f$-$5d$ spin fluctuations turns out to be substantially suppressed near the <112> directions, where the smallest values of coefficients $A_H(\varphi)$ and $B_H(\varphi)$ are observed in the $\Delta H_M$ interval. These directions correspond to the radial phase boundary of the H-$\varphi$ diagram (see Fig. 12), near which the critical fluctuations and associated inhomogeneities in the sample completely determine the carrier scattering, resulting to MR which is practically independent on magnetic field. At the end of this section, we emphasize that the complex AF state formed in $Ho_{0.8}Lu_{0.2}B_{12}$ by the combination of $4f$ (localized magnetic moments) and $5d$ (SDW) components of the magnetic structure is suppressed in strong magnetic field when the scattering of charge carriers is amplified both in nMR and pMR channels, mainly in directions along the stripes (**H** // [110]) and transverse to the stripes (**H** // [001]). Indeed, for this interval $\Delta H_H$ the coefficients $A_H(T)$ and $B_H(T)$ have the maximum values (see Figs. 15-16). At the same time, it is the scattering of conduction electrons in magnetic field **H** // [110] both on local spin fluctuations and on SDW, i.e. directed along dynamic charge stripes, that is decisive for the AF-P transition in $Ho_{0.8}Lu_{0.2}B_{12}$.

**IV.5. Root to complexity of the magnetic ground state in $Ho_{0.8}Lu_{0.2}B_{12}$.** To explain the origin of numerous magnetic phases and phase transformations in the AF state of $Ho_{0.8}Lu_{0.2}B_{12}$ as a function of temperature and external magnetic field it is necessary to take into account several important factors. Among them likely the main one which determines the complexity of the magnetic phase diagram in strongly correlated electron systems $Ho_xLu_{1-x}B_{12}$ is the emergence of dynamic charge stripes in the cage-glass state of RE dodecaborides [22-23,30]. This kind electronic instability accompanied with nanometer scale phase separation is of great importance to understand both the nature of the Maltese cross anisotropy in $Ho_xLu_{1-x}B_{12}$ antiferromagnets with a *fcc* crystal structure and the origin of local *4f-5d* fluctuations of the electron density in the nearest vicinity of $Ho^{3+}$ ions.

In this scenario, the genesis of non-equilibrium (hot) electrons in the RE dodecaborides may be explained as follows. (*i*) Because of the Jahn-Teller (JT) instability of $B_{12}$ molecule, triply degenerate electronic states of each boron cluster $[B_{12}]^{2-}$ become mixed under JT vibrations via non-adiabatic coupling [22, 43]. (*ii*) In the $RB_{12}$ matrix, the *collective* JT effect on the lattice of these $B_{12}$ complexes is at the origin of the collective dynamics of boron clusters with long-range JT disordering of the crystal lattice, which can be trigonal or tetragonal depending on the boron isotope composition [44]. As a consequence of the JT dynamics and the lattice distortions there arise large amplitude vibrations of $Ho(Lu)^{3+}$ -ions which are loosely bound with the rigid boron cage and located in a double well potential inside the $B_{24}$ truncated cubooctahedrons (see Fig.1c and [22]). (*iii*) These rattling modes of $R^{3+}$ ions necessarily initiate strong changes in the *5d–2p* hybridization of RE and boron electronic states. Since the states in the conduction band are composed by the antibonding *2p*-orbitals of $B_{12}$ molecules and the *5d* states of Ho(Lu) atoms [29,45-47], the variation of the *5d–2p* hybridization leads to the modulation of conduction band width and to consequent generation of the non-equilibrium (hot) charge carriers, the percentage of which is at room temperature between 70-80% from the total number of conduction band electrons [28]. (*iv*) In the cage-glass state of these RE dodecaborides at T< T*~ 60 K [32] two additional factors appear, 1) the positional disorder in the arrangement of $Ho^{3+}$ ions in $B_{24}$ truncated cubooctahedrons (static $Ho(Lu)^{3+}$ displacements in the double well potential) which is accompanied with formation of vibrationally coupled magnetic nanometer size clusters in the RB$_{12}$ matrix and 2) the emergence of dynamic charge stripes (*ac*-current with a frequency ~200 GHz [30]) directed along single [110] axis in Ho$_x$Lu$_{1-x}$B$_{12}$ and accumulated a considerable part of non-equilibrium conduction electrons in the filamentary structure of fluctuating charges [23]. Thus, due to the cooperative dynamic JT effect when the B$_{12}$ polyhedra are consistently distorted, both static displacements of Ho(Lu) ions and $^{10}$B-$^{11}$B substitutional disorder provide centers of pinning facilitating the formation of additional *ac* conductive channels – the dynamic charge stripes. (*v*) The positional disorder in the arrangement of Ho$^{3+}$ ions in B$_{24}$ truncated cubooctahedrons in the cage-glass state (T< T*~ 60 K) leads to a significant dispersion of exchange constants (through indirect exchange, RKKY mechanism) and formation of both nanometer size domains of magnetic Ho$^{3+}$ ions in the RB$_{12}$ matrix (short range AF order effect above T$_N$) and creation of strong local *4f-5d* spin fluctuations responsible for the polarization of *5d* conduction band states (the spin-polaron effect). The last one produces spin polarization sub-nanometer size ferromagnetic domains (ferrons, in according to the terminology used in [48-49]) resulting to stabilization of SDW antinodes in the RB$_{12}$ matrix. The spin–polarized *5d*-component of the magnetic structure (ferrons) is from one side very sensitive to the external magnetic field, and, from another side, the applied field suppresses *4f–5d* spin fluctuations by destroying the spin-flip scattering process. Moreover, along the direction of the dynamic charge stripes [110] the huge charge carrier scattering destroys the indirect exchange of Ho$^{3+}$ localized magnetic moments and renormalizes the RKKY interaction accumulating a noticeable part of charge carriers in the filamentary electronic structure. Thus, the complex *H–T-φ* phase diagrams of Ho$_x$Lu$_{1-x}$B$_{12}$ magnets may be explained in terms of the formation of a composite magnetically ordered state of localized *4f* moments of Ho$^{3+}$-ions in combination with spin polarized local areas of the *5d* states - ferrons involved in the formation of SDW in presence of filamentary structure of dynamic charge stripes. It is worth noting that the emergence of spin polarization was confirmed for HoB$_{12}$ in [33] where a ferromagnetic component of the order parameter was found above 20 kOe in magnetic neutron diffraction patterns. Moreover, even harmonics and hysteresis of Hall resistance were detected for $HoB_{12}$ in the range 20-60 kOe where ferromagnetic response was observed and these effects were attributed in [50] to charge carriers scattering on SDW.

# V. CONCLUSIONS.

A model strongly correlated electron system $Ho_{0.8}Lu_{0.2}B_{12}$ with *fcc* lattice instability (cooperative dynamic JT effect) and nanometer size electronic phase separation (dynamic charge stripes) has been studied in detail by low temperature magnetoresistance, heat capacity and magnetization measurements. The angular H-φ-T antiferromagnetic phase diagrams in the form of a Maltese cross have been deduced for this AF metal for the first time and it was matched to three main sectors in vicinity of the main directions (*i*) **H**//[110] along and (*ii*) **H**//[001] transverse to dynamic charge stripes, and (*iii*) **H**//[111] connected with the orientation of the AF magnetic structure. The same Maltese cross anisotropy was also detected for charge carriers scattering and two dominated mechanisms of the magnetoresistance in the AF state were separated and analyzed quantitatively. It was suggested that the main positive linear in magnetic field contribution may be attributed to charge carrier scattering on SDW (*5d*-component of magnetic structure), whereas the negative quadratic term $-\Delta\rho/\rho \sim H^2$ is due to scattering on local *4f-5d* spin fluctuations of $Ho^{3+}$ ions. We argue that the observed dramatic symmetry lowering is the consequence of strong renormalization of the indirect exchange interaction (RKKY mechanism) due to the presence of dynamic charge stripes in the matrix of AF metal.

# ACKNOWLEDGMENTS.

This work was supported by the Russian Science Foundation, project no. 17-12-01426 and was performed using the equipment of the Shared Facility Center for Studies of HTS and Other Strongly Correlated Materials, Lebedev Physical Institute, Russian Academy of Sciences, and of the Center of Excellence, Slovak Academy of Sciences. The work of K. Flachbart and S. Gabani was supported by the Slovak agencies VEGA (grant no. 2/0032/16) and APVV (grant no. 17-0020).

**REFERENCES**.

**FIGURE CAPTIONS.**

**Fig.1.** (a) Crystal structure of $RB_{12}$. (b) $B_{24}$ truncated cubooctahedrons surrounding two adjacent $R^{3+}$ ions. (c) Schematic view of R ions vibrations in the double-well potentials. Red arrow shows the direction [110] of dynamic charge stripes in $Ho_xLu_{1-x}B_{12}$. (d) Schematic view on measurements of the magnetoresistance angular dependence. **n**- normal vector to the surface of the studied crystal, **H**-magnetic field, **I**- measuring current, φ- angle between **n** and **H** vectors.

**Fig. 2.** Temperature dependences of resistivity, specific heat and magnetic susceptibility (H = 100 Oe) in the wide vicinity of Neel temperature for $Ho_{0.8}Lu_{0.2}B_{12}$.

**Fig.3.** Magnetic field dependences of magnetoresistance for (a) **H**//[001], (b) **H**//[110] and **H**//[111] at temperatures in the range of 1.9-6.6 K. Arrows indicate the magnetic phase transitions.

**Fig.4.** Temperature dependences of specific heat for (a) **H**//[001], (b) **H**//[110] and (c) **H**//[111] in magnetic field up to 70 kOe. Arrows indicate the magnetic phase transitions.

**Fig.5.** Magnetic field dependences of magnetization per Ho ion (panels a-c) and their derivatives (panels d-f) for (a), (d) **H**//[001], (b), (e) **H**//[110] and (c), (f) **H**//[111] at temperatures in the range 1.9-6.6 K. Arrows indicate the antiferromagnetic-paramagnetic phase transitions.

**Fig.6.** Magnetic field dependences of dM/dH derivatives for H< 25 kOe along directions (a) **H**//[001], (b) **H**//[110] and (c) **H**//[111] at temperatures in the range 1.9-5.7 K. Arrows indicate the orientation magnetic phase transitions.

**Fig.7.** Magnetic field dependences of magnetization derivatives dM/dH (panels a, d, g), specific heat (panels b, e, h) and magnetoresistance derivatives (panels c, f, i) along directions **H**//[001] (a-c), **H**//[110] (d-f) and **H**//[111] (g-i) at 2.1 K. Arrows, symbols and vertical dotted lines indicate magnetic phase transitions. Panels (c, f, i) show also the linear approximation of the magnetoresistance derivative. Inset in panel (c) demonstrates an additional phase transition at H=52 kOe.

**Fig. 8.** Magnetic field dependences of magnetoresistance at T=2.1 K for different directions of the external magnetic field in the range (a) between **H**//[001] and **H**//[111] and (b) between **H**//[111] and **H**//[110]. Arrows and symbols indicate magnetic phase transitions.

**Fig.9.** Angular dependences of resistivity at T = 2.1 K and in various magnetic fields up to 80 kOe. Top axis shows the principal axes.

**Fig. 10.** H-T magnetic phase diagram of $Ho_{0.8}Lu_{0.2}B_{12}$ for magnetic fields applied in the three principal cubic directions. The phase boundaries are derived from magnetoresistance, heat capacity and magnetization data (see symbols in panel (b)). Roman numerals denote various phases within the antiferromagnetic state. P- paramagnetic phase. Red vertical dotted lines indicate three temperatures (T = 2.1 K, 4.2 K and 6.5 K) at which the resistivity angular dependences have been studied in detail.

**Fig.11.** 3D view of magnetoresistance dependences on magnetic field (a) T = 2.1 K, (b) T = 4.2 K and (d) T = 6.5 K, and on temperature (c) at H = 25 kOe. The rotation axis is **I**//[1-10], the three principal cubic directions in the (1-10) plane are shown by arrows.

**Fig.12.** H-φ magnetic phase diagram of $Ho_{0.8}Lu_{0.2}B_{12}$ at temperatures 2.1 K (a), 4.2 K (b) and T-φ diagram in magnetic field H = 25 kOe. The phase boundaries are derived from magnetoresistance data (see symbols in fig. 8, for example). Roman numerals are the same as in fig. 10 and denote various phases within the antiferromagnetic state.

**Fig. 13.** Scaling of magnetoresistance $\Delta\rho/\rho=f(H/T)$ in the AF state for magnetic fields applied in the three principal cubic directions at temperatures in the range 1.9-6.6 K. The slope of the linear dependence is shown by coefficient $k$.

**Fig. 14.** Magnetic field dependences of resistivity derivatives at temperatures in the range 1.9-4.2 K for magnetic field directions (a) **H**//[001] and (b) **H**//[111]. Solid lines show the linear approximation (see Eq.(3) in the text) in intervals $\Delta H_L$, $\Delta H_M$ and $\Delta H_H$. Curves are vertically shifted for convenience.

**Fig.15.** Temperature dependences of the coefficients (a) $A_j(T)$ and (b) $B_j(T)$ (j=L, M and H) deduced from the analysis of magnetoresistance components within the framework of Eq. (3) for ~~the~~ magnetic fields applied in the three principal cubic directions. Vertical dotted lines correspond to changes of the regimes of charge carrier scattering (see text for details).

**Fig.16.** Angular dependences of the coefficients (a-c) $A_j(T)$ and (b-d) $B_j(T)$ (j=L, M and H) deduced from the analysis of magnetoresistance components within the framework of Eq. (3) at T = 2.1 K (panels a-b) and 4.2 K (panels c-d, see text for details).

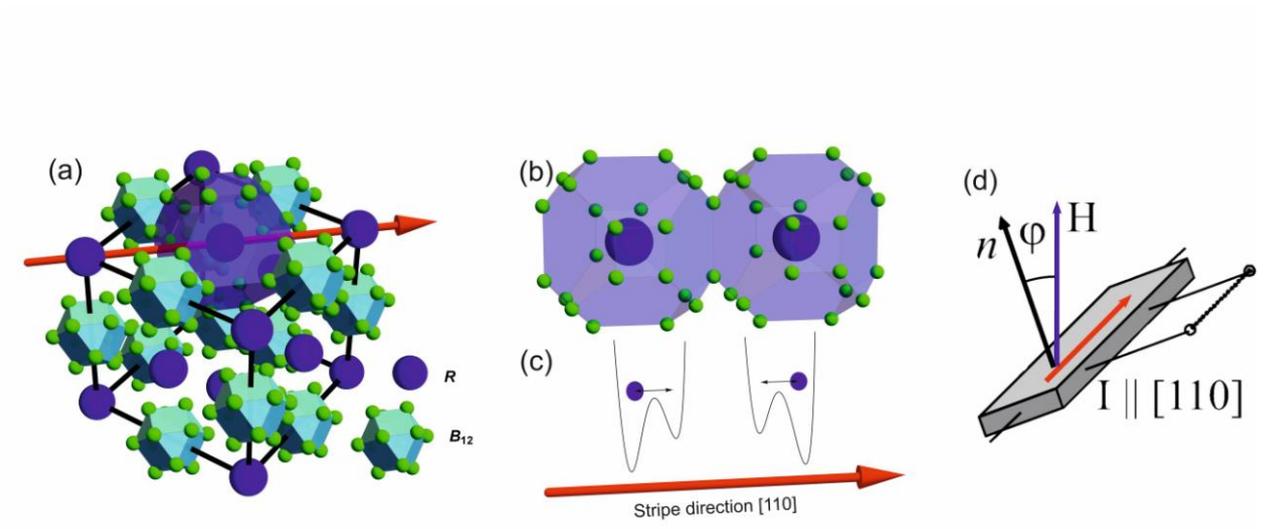

Fig.1.

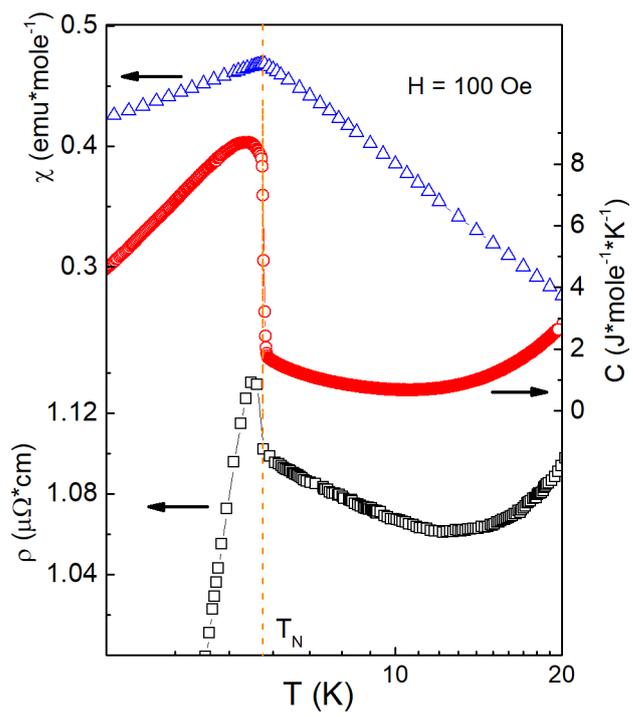

Fig.2.

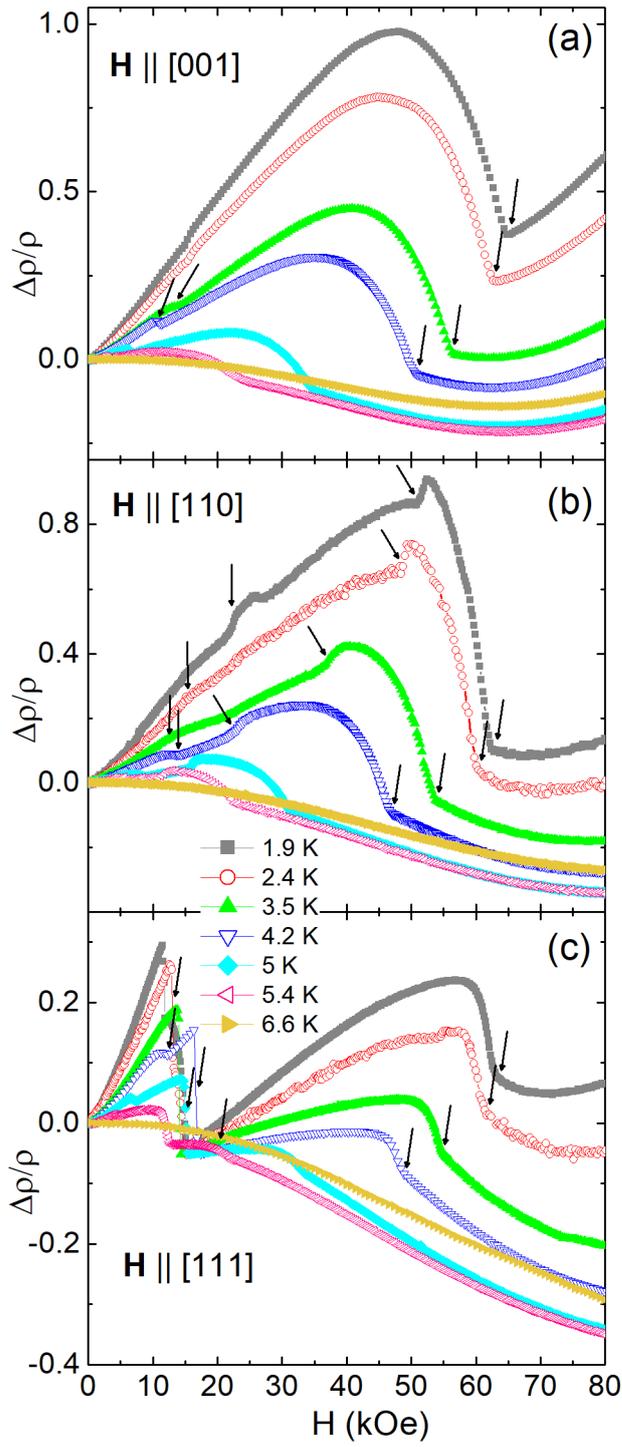

Fig.3.

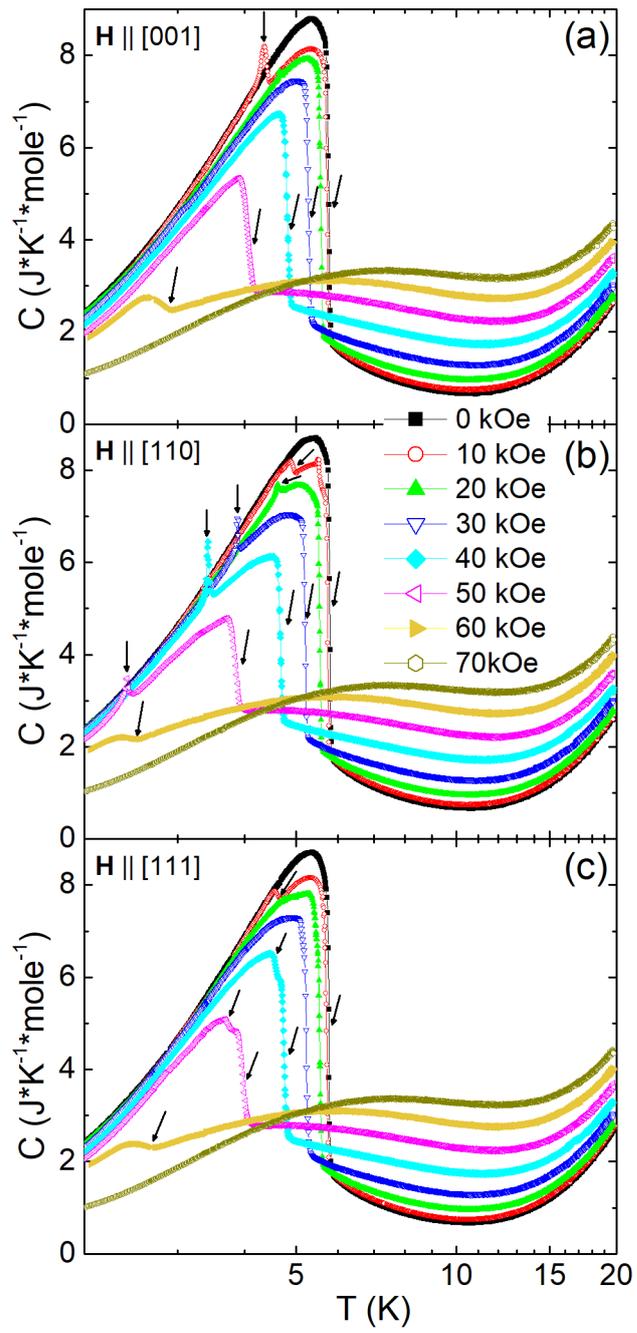

Fig.4.

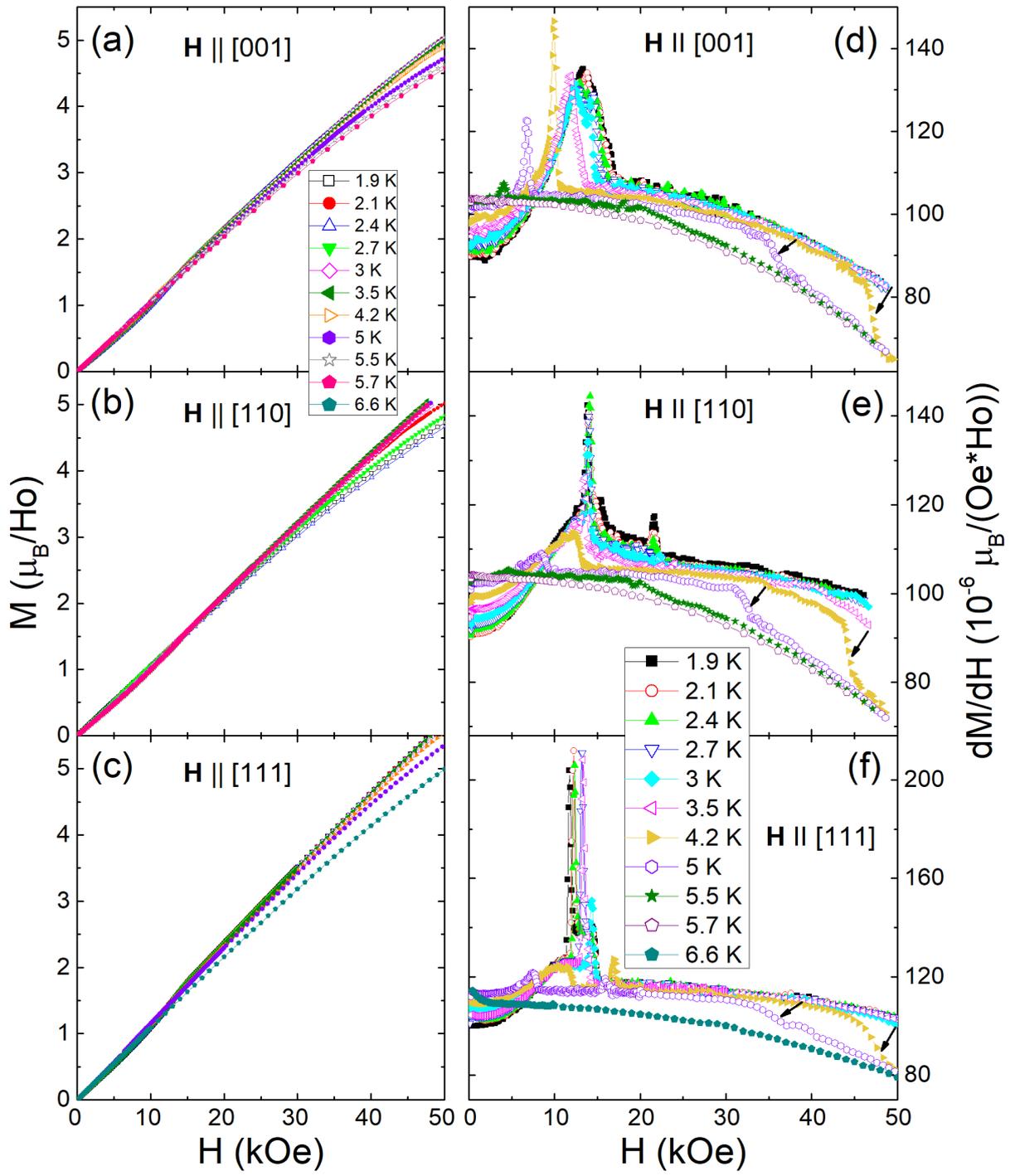

Fig.5.

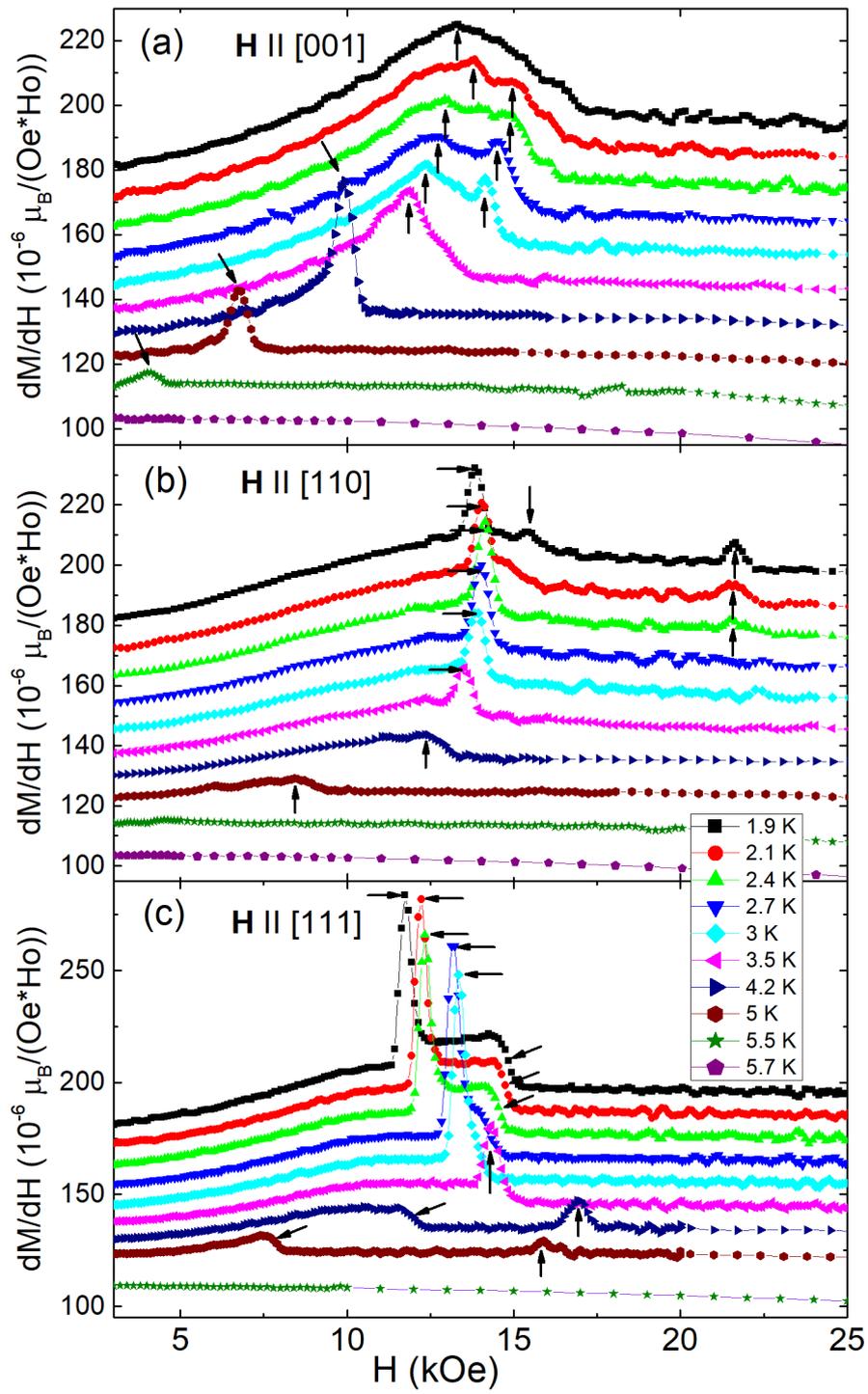

Fig.6.

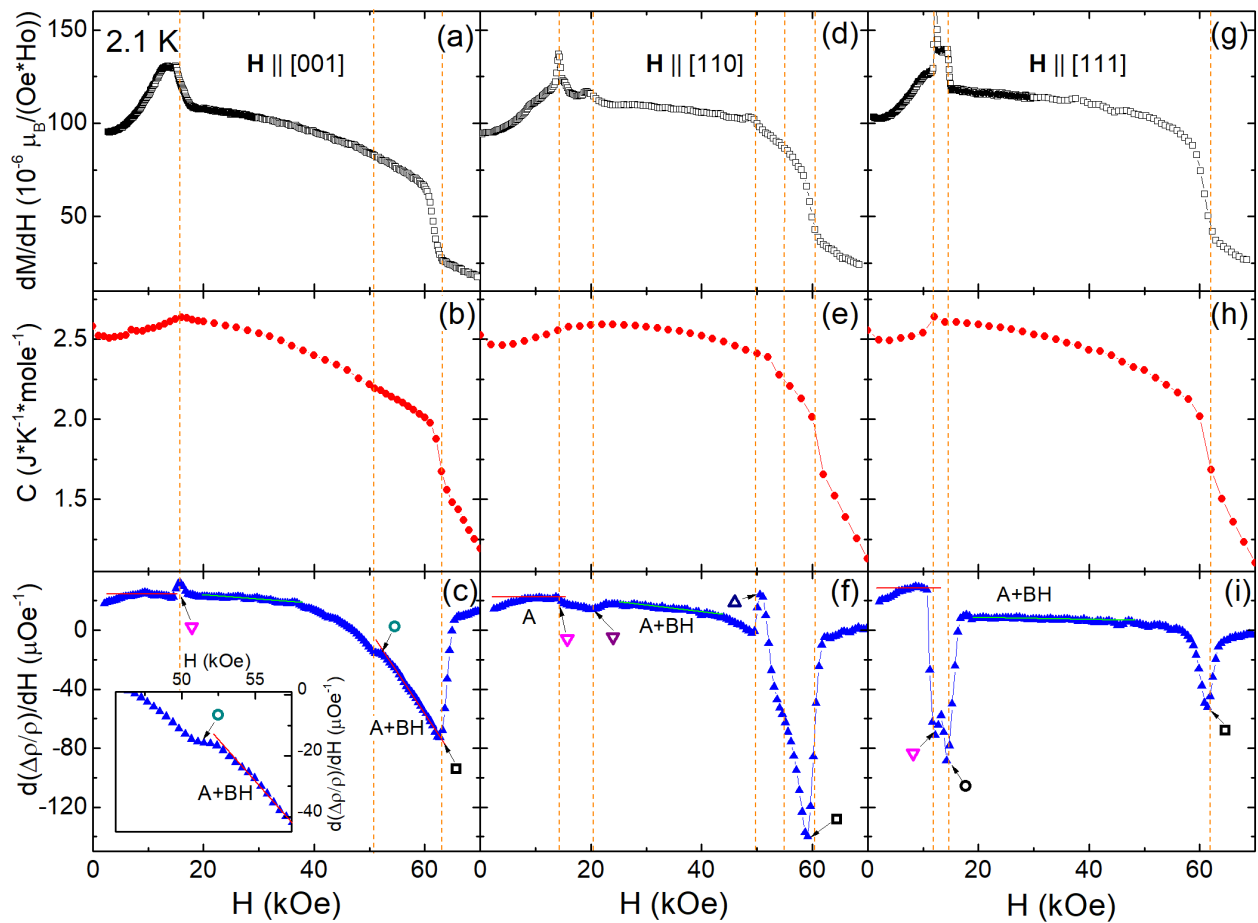

Fig.7.

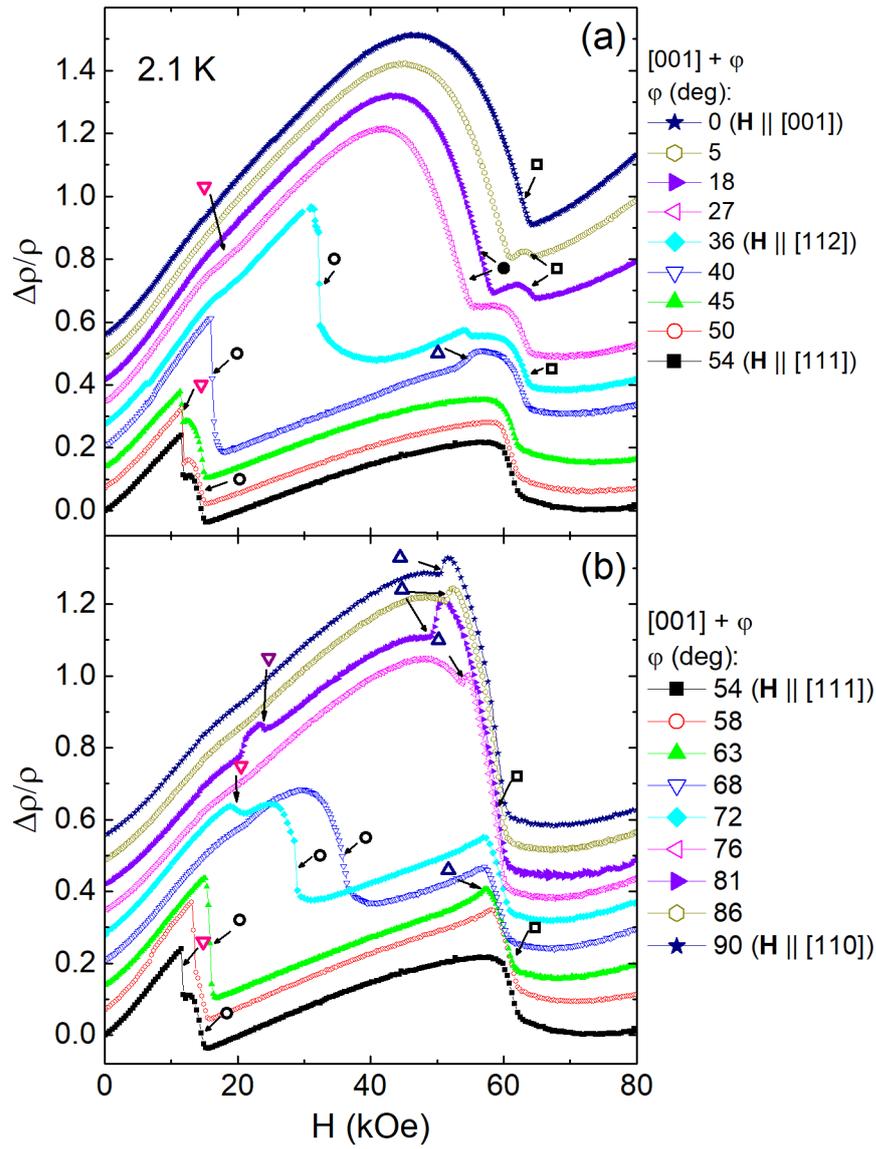

Fig.8.

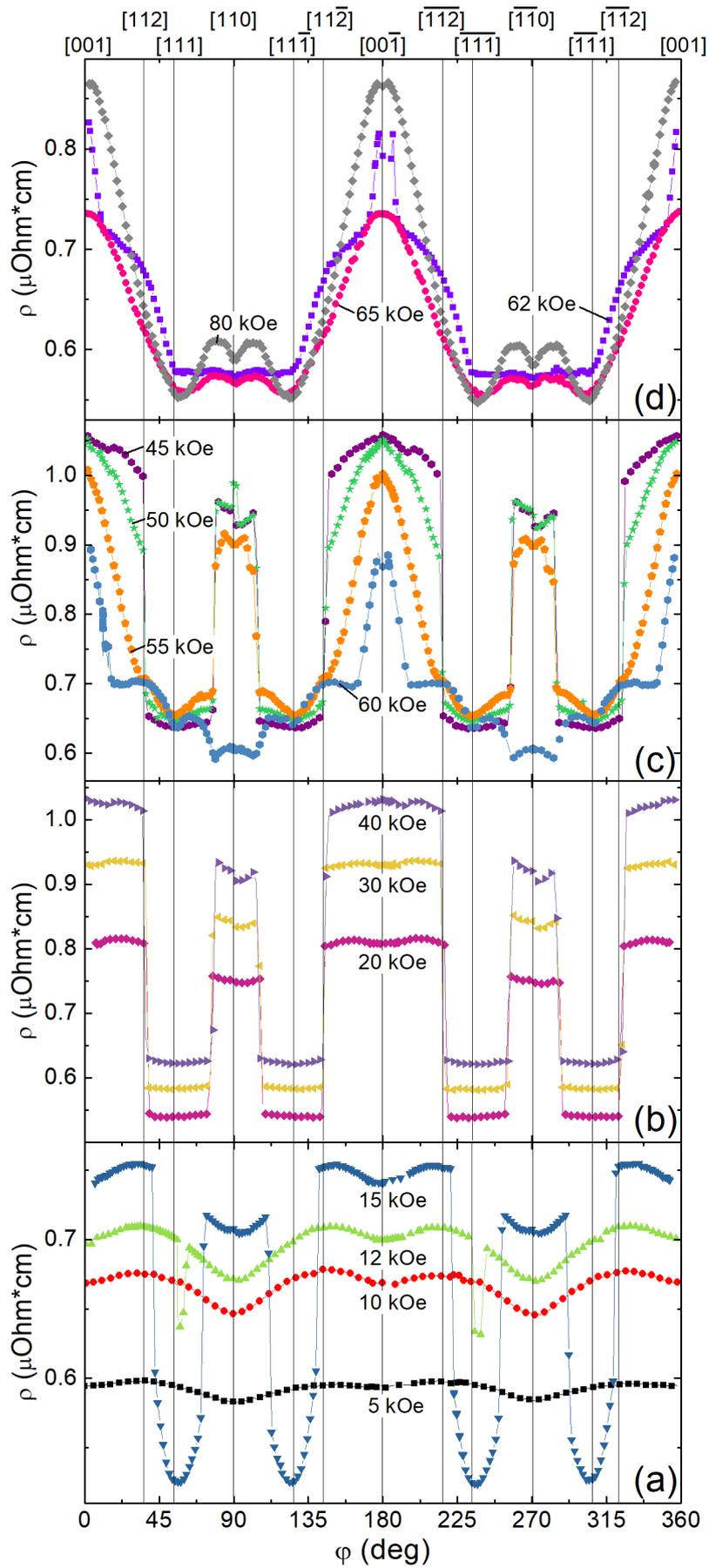

Fig.9.

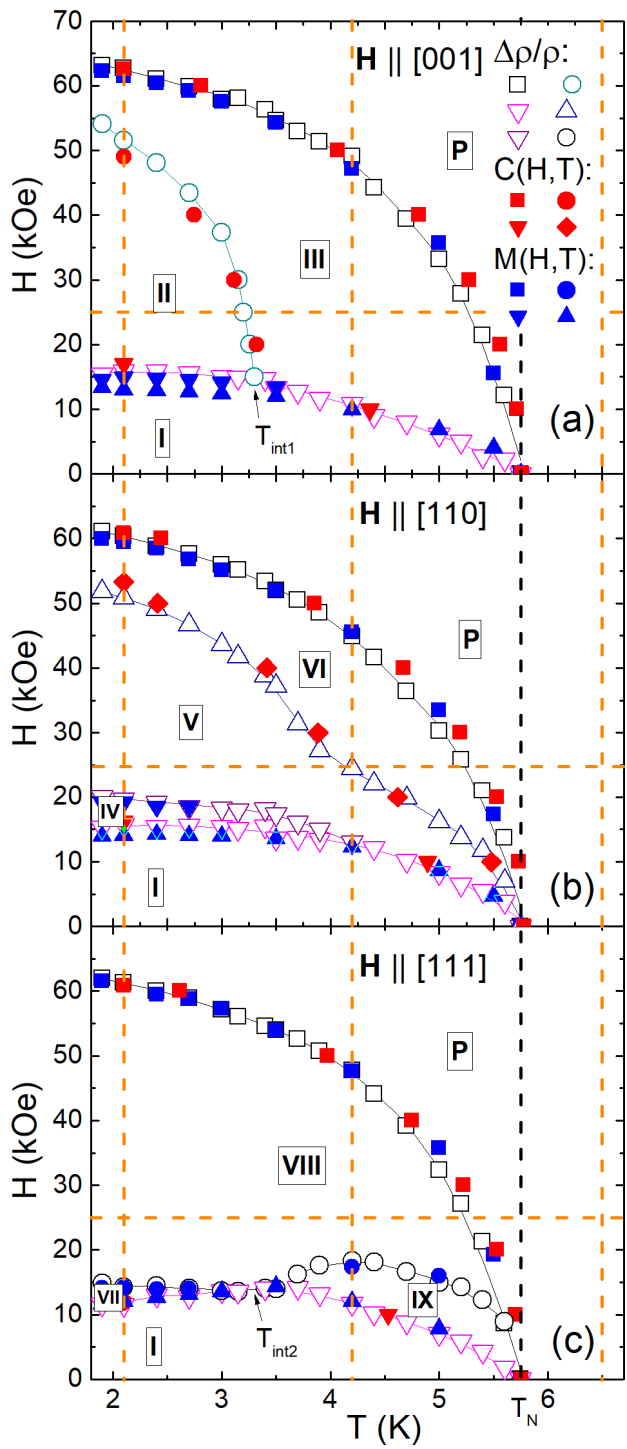

Fig.10.

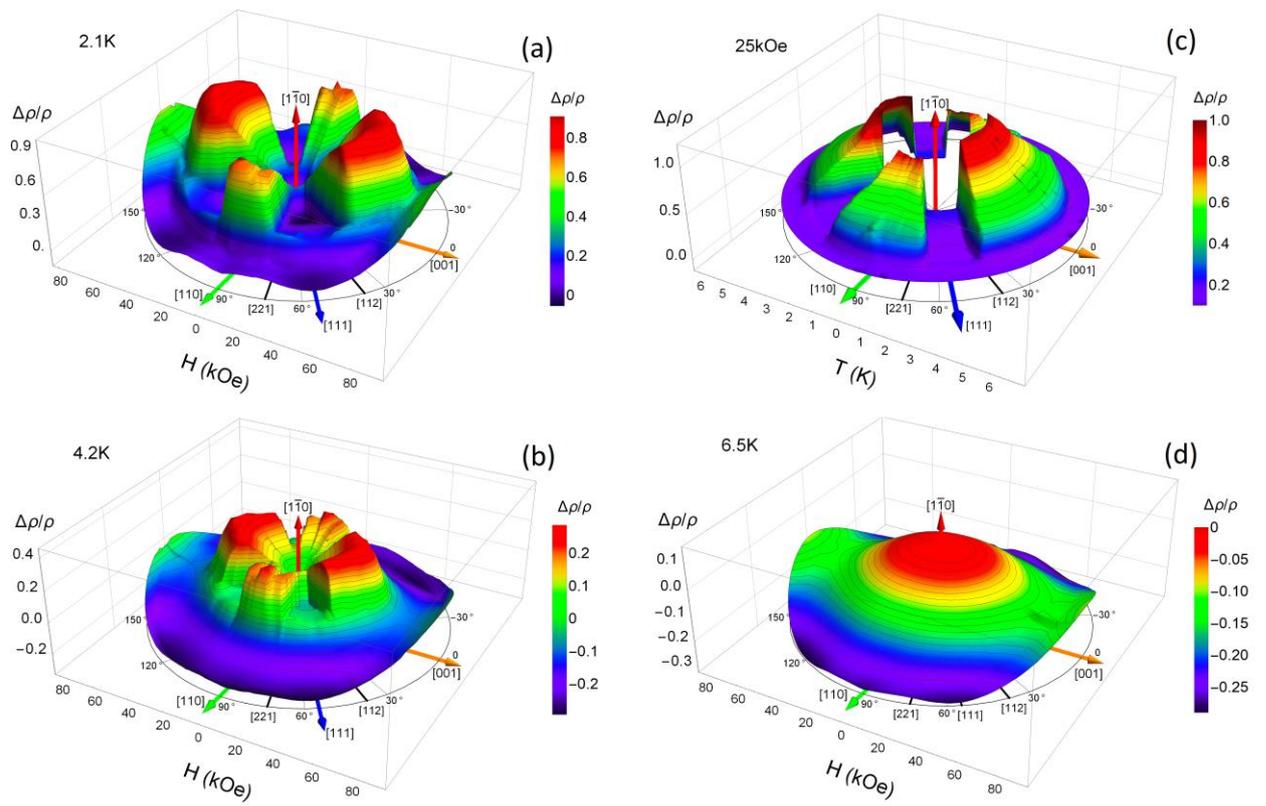

Fig.11.

Fig.12.

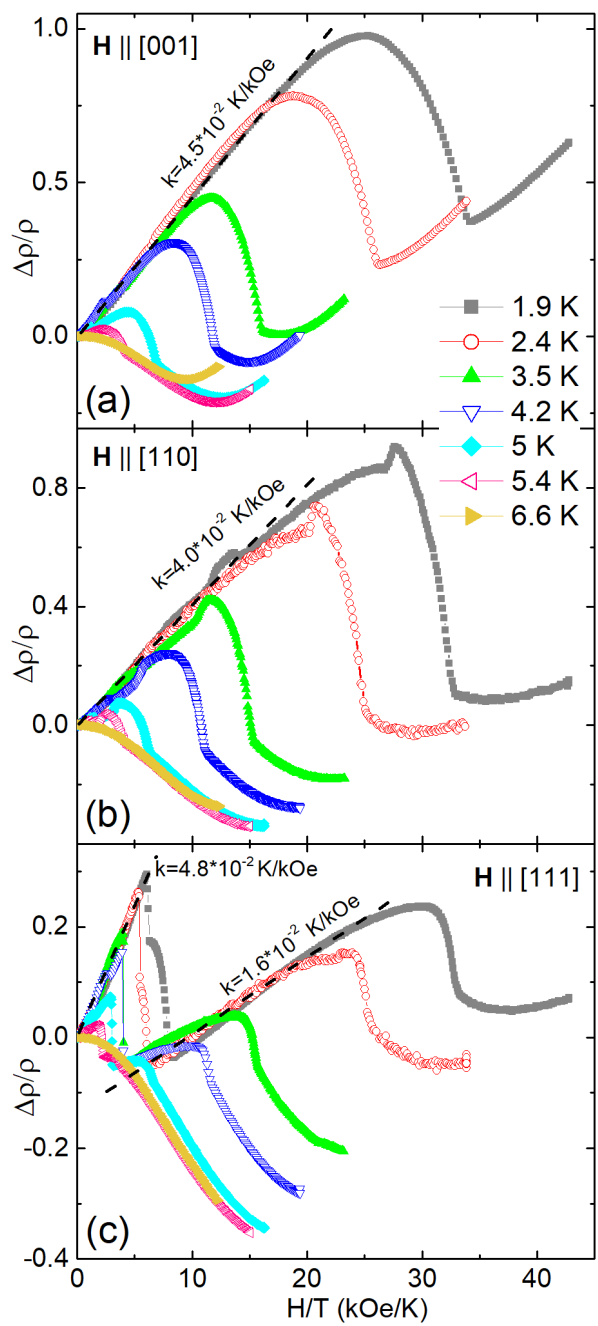

Fig.13.

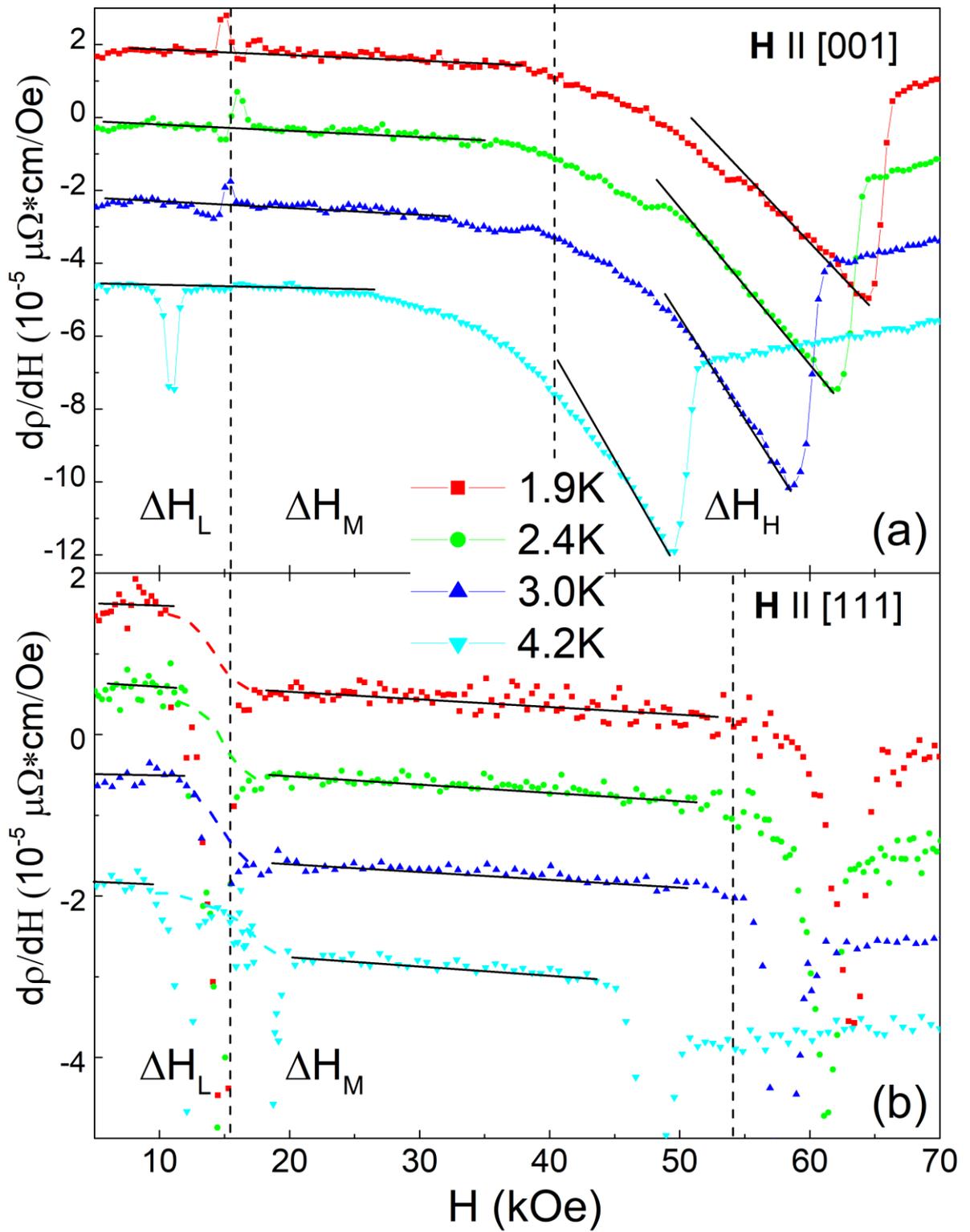

Fig.14.

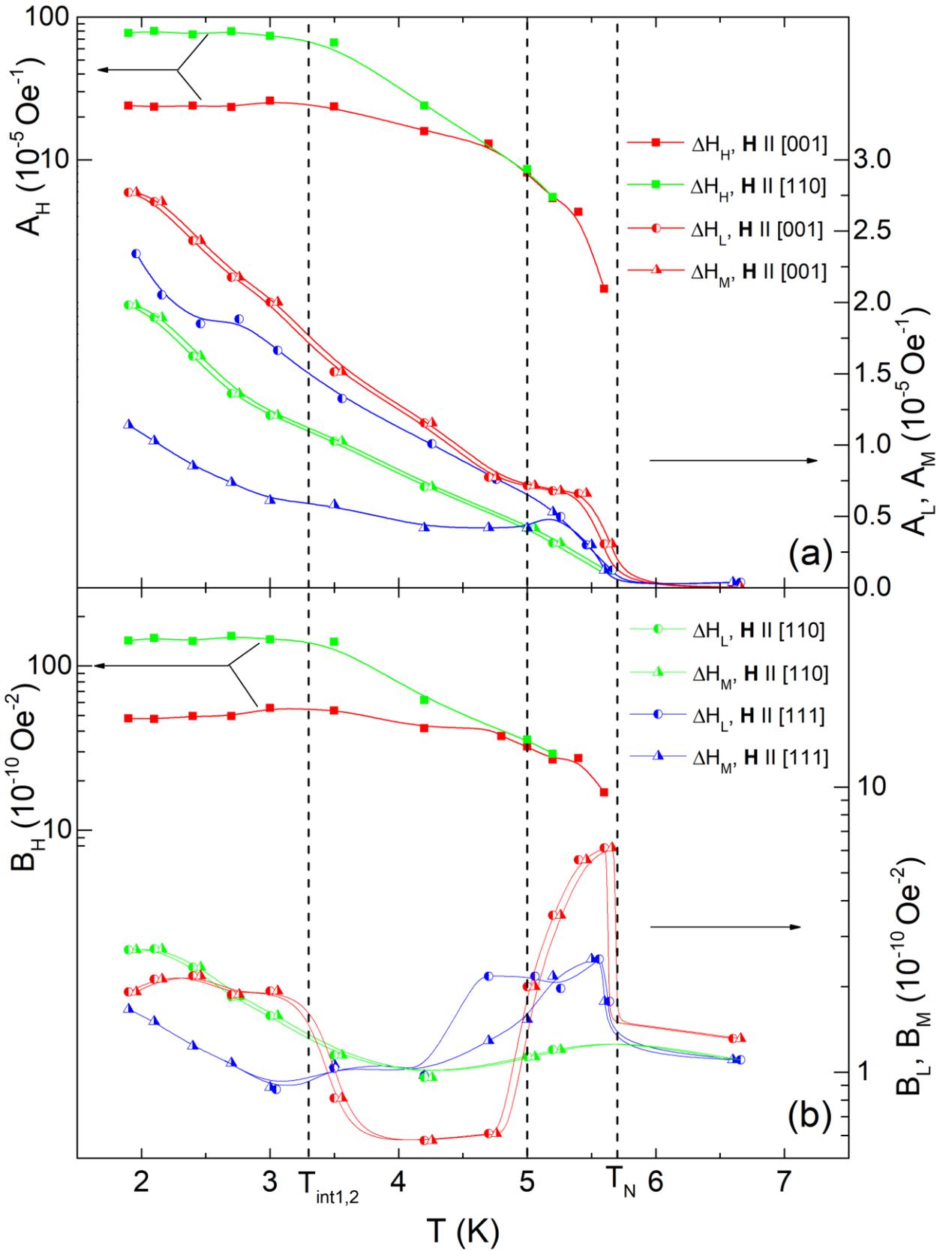

Fig.15.

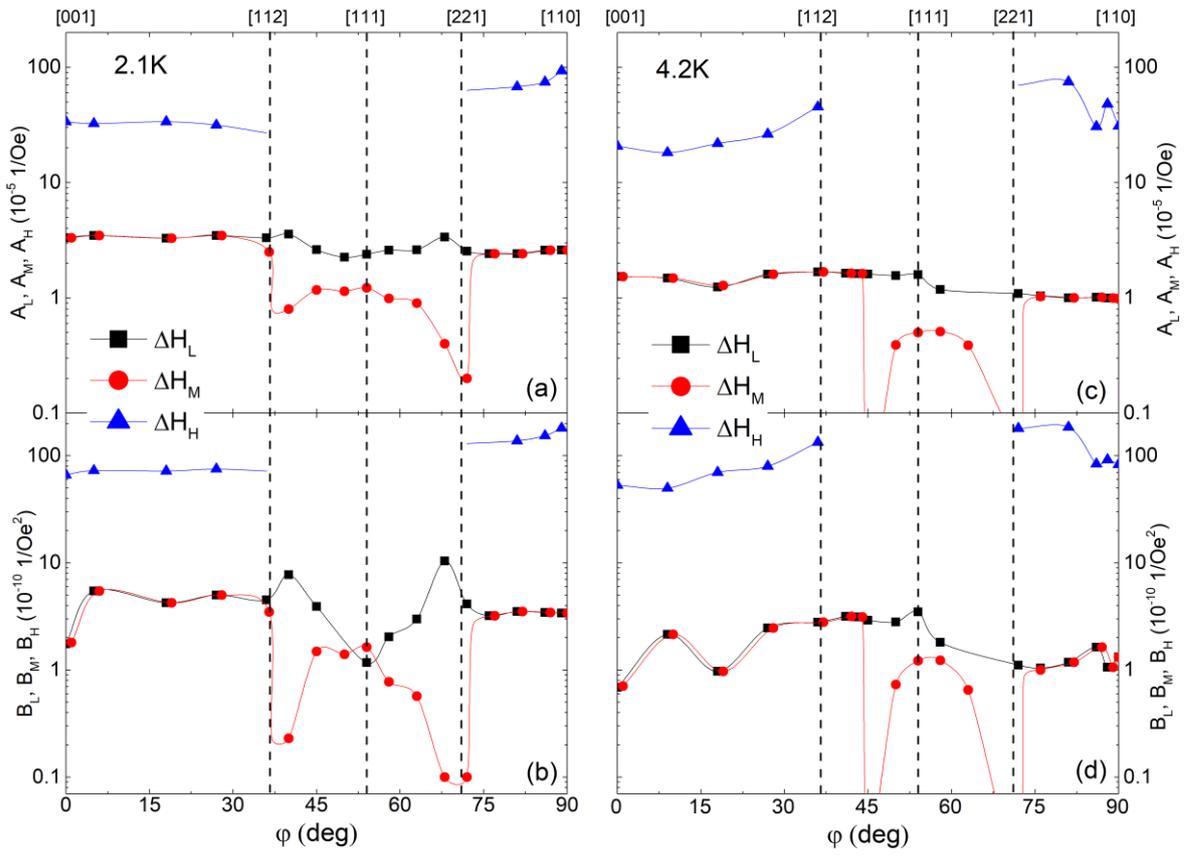

Fig.16.